\documentclass[12pt]{article}
\usepackage{epsfig,amsfonts,amsthm}
\newcommand{\be}{\begin{equation}}
\newcommand{\ee}{\end{equation}}
\newcommand{\bea}{\begin{eqnarray}}
\newcommand{\eea}{\end{eqnarray}}

\newcommand{\br}{{\bf r }}

\newcommand{\bk}{{\bf k}}

\newcommand{\bn}{{\bf n}}
\newcommand{\bp}{{\bf p}}
\newcommand{\bq}{{\bf q}}

\def\lsim{\mathrel{\rlap{\lower4pt\hbox{\hskip1pt$\sim$}}
    \raise1pt\hbox{$<$}}}         

\def\gsim{\mathrel{\rlap{\lower4pt\hbox{\hskip1pt$\sim$}}
    \raise1pt\hbox{$>$}}}         

\title{Colliding particles carrying non-zero orbital angular momentum}
\author{Igor~P.~Ivanov
\\
  {\small IFPA, Universit\'{e} de Li\`{e}ge, All\'{e}e du 6 Ao\^{u}t 17, b\^{a}timent B5a, 4000 Li\`{e}ge, Belgium}\\
  {\small and}\\
  {\small Sobolev Institute of Mathematics, Koptyug avenue 4, 630090, Novosibirsk, Russia}\\
  }

\begin{document}

\maketitle

\begin{abstract}
Photons carrying non-zero orbital angular momentum (twisted photons)
are well-known in optics. Recently, it was suggested to use Compton backscattering
to boost optical twisted photons to high energies.
Twisted electrons in the intermediate energy range have also been produced recently.
Thus, collisions involving energetic twisted particles seem to be feasible and
represent a new tool in high-energy physics.
Here we discuss some generic features of scattering processes involving twisted particles in the initial
and/or final state. In order to avoid additional complications arising from non-trivial polarization
states, we focus here on scalar fields only.
We show that processes involving twisted particles allow one to perform 
a Fourier analysis of the plane wave cross section 
with respect to the azimuthal angles of the initial particles.
In addition, using twisted states one can probe the autocorrelation function of the amplitude, 
which is inaccessible in the plane wave collisions.
Finally, we discuss prospects for experimental study of these effects.
\end{abstract}

\section{Introduction}

In perturbative quantum field theory we assume that interaction among the fields
can be treated as a perturbation of the free field theory.
This perturbation leads to scattering between asymptotically free multiparticle states,
which are usually constructed from the plane wave one-particle states.
This choice greatly simplifies the calculations
and represents a very accurate approximation to the real experimental situation
in virtually all circumstances.
However, one can, in principle, choose any complete basis 
for the one-particle states other than the plane wave basis, provided
that it is still made up of solutions of the free field equations.
Such states can carry new quantum numbers absent in the plane wave choice
and, if experimentally realized, they can offer new opportunities in high-energy physics.

Thanks to the progress in optics made in the last two decades, it is now possible to create 
laser beams carrying non-zero orbital angular momentum (OAM)
\cite{OAM}, for a recent review see \cite{OAMreview}.
The lightfield in such beams is described via non-plane wave solutions of the Maxwell equations. 
Each photon in this lightfield, which we call a {\em twisted photon}, 
carries a non-zero OAM quantized in units of $\hbar$.
Several sets of solutions have been investigated, such as Bessel beams or Gauss-Laguerre beams,
but in all cases the spatial distribution of the lightfield is necessarily 
non-homogeneous in the sense that the equal phase fronts are not planes but helices.
Such states form a complete basis which can be used to describe the initial and final 
asymptotically free states. Moreover, it is the basis of choice for experimental situations 
when the initial states are prepared in a state of (more or less) definite OAM.

Twisted photons have been produced in various wavelength domains, from radiowave \cite{radio} 
to optical, with prospects to create a brilliant X-ray beam of twisted light in the keV range 
\cite{Xrayproposal}.
Very recently it was suggested to use the Compton backscattering
of twisted optical photons off an ultra-relativistic electron beam to create
a beam of high-energy photons with non-zero OAM \cite{serbo1,serbo2}.
The technology of Compton backscattering is well established  \cite{backscattering}, and the 
high-energy electron beams and the OAM optical laser beams are already available.
In addition, in the last months several groups have reported successful creation of 
twisted electrons, first using phase plates \cite{twisted-electron} and then with
computer-generated holograms \cite{twisted-electron2}.
Twisted electrons carried the energy as high as 300 keV and the orbital quantum number up to $\sim 100$.
With all these achievements, creating high-energy particles in a 
controlled orbital angular momentum state and colliding them seems now feasible.  
It is therefore very timely to ask what new insights into the properties of particles and their interactions 
one can gain with this new degree of freedom.

In this paper we begin this exploration by studying several generic scattering processes involving twisted particles
in the initial and/or final states. Namely, we consider three specific cases:
\begin{itemize}
\item
single-twisted scattering: collision of a twisted state with a plane wave,
\item
double-twisted scattering: collision of two twisted states,
\item
two-particle decay of an unstable twisted particle.
\end{itemize}
In the first two cases we assume that the final system $X$ is described by plane waves,
while in the last case we consider three choices for the final two-particle state: 
when both particles are plane waves, when one is twisted, and when both are twisted.
The single- and double-twisted scattering will give some hints at new physical opportunities that can be expected
in collisions of high-energy particles carrying OAM, while the calculation of a twisted particle decay clarifies 
various technical details involved in passage from plane waves to twisted states.

The OAM and spin are two forms of angular momentum,
and the problem of gauge-invariant separation of these two objects has a long history.
It is still being debated both in the optics community, see for example \cite{bliokh}, 
and in the HEP community, especially in the context of the notorious proton spin puzzle, \cite{proton-spin}.
For the problems we consider here it is sufficient to note that all experimental situations
which seem to be realizable are well described by the paraxial approximation.
It is known that in the paraxial approximation spin can be well separated 
from the $z$-component of OAM of light, \cite{spinOAMseparation}. 
The same applies to fermions as well.
Therefore, incorporation of both spin and OAM degrees of freedom,
leading to non-trivial polarization fields, does not seem to pose any problem
in the paraxial approximation. 

However in the present paper we would like to focus specifically on the OAM component
and to understand what new physical opportunities are offered by the non-trivial spatial dependence rather
than unusual polarization states. 
Therefore we limit ourselves to scattering of {\em scalar} particles only.
In this simple case all the non-zero angular momentum is definitely due to the orbital part.
Additional features arising from the polarization parameter fields
will be considered separately.

The paper is organized as follows.
In Section \ref{section-twisted} we introduce scalar twisted states and 
describe some of their properties.
In Sections \ref{section-single} and \ref{section-double} we derive expressions
for the cross section in the single-twisted and double-twisted cases, respectively.
Section \ref{section-decay} gives a thorough discussion of kinematical features arising in the case when
the final state contains twisted particles. 
In Section \ref{section-discussion} we discuss the results obtained
and draw our conclusions. In two Appendices we derive some technical results
used in the paper.

\section{Describing twisted states}\label{section-twisted}

\subsection{Spatial distribution}

As mentioned in the introduction, we focus in this paper 
on twisted scalar particles with mass $M$. In their description we follow essentially
\cite{serbo1,serbo2}.

We represent a state with a non-zero OAM with a Bessel beam-type twisted state.
This is a solution of the wave equation in the cylindric coordinates with a definite
energy $\omega$ and a longitudinal momentum $k_z$ along a fixed axis $z$, 
a definite modulus of the transverse momentum $|\bk|$ (all transverse momenta will be written in bold)
and a definite $z$-projection of OAM.
If the plane wave state $|PW(\bk)\rangle$ is
\be
|PW(\bk)\rangle = e^{-i\omega t + i k_z z} \cdot e^{i\bk\br}\,,
\ee
then a twisted scalar state $|\kappa,m\rangle$ is defined as the following superposition
of plane waves: 
\be
|\kappa,m\rangle = e^{-i\omega t + i k_z z} 
\int {d^2\bk \over(2\pi)^2}a_{\kappa m}(\bk) e^{i\bk \br}\,,
\quad a_{\kappa m}(\bk)= (-i)^m e^{im\phi_k}\sqrt{2\pi}{\delta(|\bk|-\kappa)\over \sqrt{\kappa}}\,.
\label{twisted-def}
\ee
In the coordinate space,
\be
|\kappa, m\rangle = e^{-i\omega t + i k_z z} \cdot \psi_{\kappa m}(\br)\,,
\quad \psi_{\kappa m}(\br) = {e^{i m \phi_r} \over\sqrt{2\pi}}\sqrt{\kappa}J_{m}(\kappa r)\,.
\ee
Here, following \cite{serbo1} we call $\kappa$ the conical momentum spread, $m$ is the $z$-projection of OAM,
and the dispersion relation is $k^\mu k_\mu = \omega^2 - k_z^2 - \kappa^2 = M^2$.
We note in passing that the average values of the four-momentum carried by a twisted state is
\be
\langle k^\mu \rangle = (\omega,\, {\bf 0},\, k_z)\,,
\ee
so that $\langle k^\mu \rangle \langle k_\mu \rangle = M^2 + \kappa^2$, which is larger than the true mass of the particle squared.

The transverse spatial distribution is normalized according to
\be
\int d^2\br \psi^*_{\kappa' m'}(\br)\psi_{\kappa m}(\br) = 
\delta_{m,m'} \sqrt{\kappa\kappa'}\int rdr J_{m}(\kappa r) J_{m}(\kappa' r) = \delta_{m,m'}\delta(\kappa-\kappa')\,.
\ee
The plane wave can be recovered from the twisted states as follows:
\bea
|PW(\bk = 0)\rangle & = & \lim_{\kappa \to 0} \sqrt{{2\pi \over\kappa}} |\kappa,0\rangle\,,\label{PWlimit0}\\
|PW(\bk)\rangle & = & \sqrt{{2\pi \over\kappa}} 
\sum_{m=-\infty}^{+\infty} i^m e^{-im\phi_k} |\kappa,m\rangle \,,\quad 
\kappa = |\bk_\perp|\,.\label{PWlimit}
\eea
If needed, these two cases can be written as a single expression:
\be
|PW(\bk)\rangle = \lim_{\kappa \to |\bk|} \sqrt{{2\pi \over\kappa}}
\sum_{m=-\infty}^{+\infty} i^m e^{-im\phi_k} |\kappa,m\rangle\,.
\ee
From these expressions one sees that the twisted states with different $m$ and $\kappa$
represent nothing but another basis for the transverse wave functions.

\subsection{Density of states}

When calculating cross sections and decay rates, we need to integrate the transition
probability over the phase space of the final particles.
When calculating the density of states, we consider a large but finite volume 
and count how many mutually orthogonal states with prescribed boundary conditions
can be squeezed inside. In the present case due to the cylindrical symmetry of the problem, 
we choose a cylinder of a large radius $R$ and a length $L_z$.
In the case of plane waves we have
\be
dn_{PW} = \pi R^2 L_z{dk_z d^2\bk \over (2\pi)^3}\,.
\ee
The full number of states with transverse momenta up to $\kappa_0$ and longitudinal momenta
$|k_z| \le k_{z0}$ is $k_{z0} L_z\cdot R^2\kappa_0^2/4\pi$.

To count the number of twisted states $|\kappa,m\rangle$ in the same volume, 
we specify the boundary condition, e.g. $\psi_{\kappa m}(r=R)=0$, which makes
$\kappa$ discrete such that $\kappa_i R$ is the $i$-th
root of the Bessel function $J_m$. 
Then we note that the position of the first root of the Bessel function $J_m(x)$
is always at $x > m$, and as $m$ grows $x \to m$.
For a given $\kappa$, the maximal $m$ for which the wave can still be contained
inside the cylindrical volume is $m_{max} = \kappa R$,
which has a very natural quasiclassical interpretation. 

If $m$ is small and not growing with $R$, then one can use the well-known
asymptotic form of the Bessel functions to count the number of states:
\be
dn_{tw} = {R d\kappa \, L_z dk_z\, \Delta m\over 2\pi^2}\,.\label{dn_tw1}
\ee
Here, $\Delta m$ is written instead of just 1 to signal the presence of 
a discrete running parameter $m$.

If $m$ is not restricted to small values, this asymptotic form 
of $J_{m}(x)$ cannot be used since it requires $m^2 \lsim x$. 
Instead, the so-called approximation by tangents can be used, 
which gives the following density of states:
\be
dn_{tw} = \sqrt{m_{max}^2-m^2} {d\kappa \over \kappa}{\Delta m \over \pi}{L_z dk_z \over 2\pi}\,.
\ee
In the limit $m \ll m_{max}\equiv \kappa R$ this expression reproduced (\ref{dn_tw1}).
Alternatively, one can calculate the radial part of the density of states
via the adiabatic invariant as suggested in \cite{serbo2}.
The number of radial excitations $n_r$ for a fixed $m$ is
\be
n_r = \int_{m/\kappa}^{R}{k_r(r)dr \over \pi}\,,\quad k_r(r) = \sqrt{\kappa^2 - {m^2\over r^2}}\,. 
\ee
The density of states is then given by
\be
dn_r = {dn_r \over d\kappa} d\kappa = \sqrt{m^2_{max}-m^2}{d\kappa \over \kappa \pi}\,.
\ee

One important remark is in order.
Effectively, switching from the plane wave to twisted state basis for the final particles
implies replacement
\be
d^2\bk \to 4\sqrt{1-{m^2 \over m_{max}^2}}\, \kappa d\kappa\, {\Delta m \over m_{max}}\,.
\ee
Note that the contribution of each ``partial wave'' with a fixed $m$ vanishes in the infinite volume limit as $1/R$.
However, the number of partial waves grows $\propto R$, and in order to get a non-vanishing
result for a physical observable, one must integrate over the full available $m$ interval up to $m_{max}$.
This remains true even if the transverse momenta stay small, and it is related to the fact that 
the plane wave contains contributions from all impact parameters with respect to any axis non-collinear
to its propagation direction.

Another expression one needs for the probability calculations is the
normalization constants for the one-particle states. 
A usual plane wave one-particle state is normalized to $2E\cdot V$;
to renormalize it to one particle per the entire volume, the plane wave
should be multiplied by $N_{PW}$, with 
\be
N^2_{PW} = {1 \over 2E V}\,,\quad V = \pi R^2 L_z\,.\label{norm-PW}
\ee
For a twisted state the corresponding normalization factor $N_{tw}$ is
\be
N_{tw}^2 = {1 \over 2E}{\pi \kappa\over \sqrt{m_{max}^2-m^2} L_z}\,,\label{norm-tw}
\ee
which in the small-$m$ case simplifies to 
\be
N_{tw}^2 \approx {1 \over 2E} {\pi \over R L_z}\,,\label{norm-tw2}
\ee
also derived in \cite{serbo2}.
Note however that even in the general case
the product of the normalization constant squared and the density of states for each final
twisted particle is simplified as
\be
N_{tw}^2 dn_{tw} = {d\kappa dk_z \Delta m \over 2E \cdot 2\pi}\,.
\ee

\subsection{Flux and the cross section}\label{subsection-flux}

The definitions of the flux factor and the cross section have to be reevaluated when 
a collision of non-plane wave states is considered, which involves subtle issues described in \cite{serbo2}.
By definition, the cross section is the transition probability per unit time
divided by flux. In the plane wave case, when the four-momenta of colliding particles
are fixed, both the flux and probability are constant
across any chosen plane, so the proportionality between them holds locally.
For a head-on collision one gets $j_{PW} = (|v_1|+|v_2|)/V$, which together with
the energies of the incoming particles and one volume factor combine to the 
familiar Lorentz invariant expression
\be
I_{PW} = \sqrt{(pk)^2-p^2k^2}\,.
\ee
This formula can of course be used in any frame, including cases when the collision is not
head-on. 

In the single-twisted case, when a twisted state collides with a plane wave,
both the flux and the transition probability are not constant but change across the transverse plane.
As noted in \cite{serbo2}, one therefore needs to redefine the notion of the cross section 
to adapt to this situation.
One introduces the {\em averaged cross section} 
defined as the transition probability integrated over all $\br$ divided
by the flux again integrated over all $\br$.
While the definition of the former quantity is clear (this is what we calculate 
in the following two Sections), the proper definition of the integrated flux is more intricate
and apparently not unique. A definition of the flux should however be correlated
with a definition of the (generalized) luminosity, so that the observable event rate remains uniquely defined.

In \cite{serbo2} the forward electron-photon collision was considered
(the 3-momentum of the plane wave electron was directed exactly along the axis $z$),
and the following procedure was suggested: the total flux is just sum of the $z$-components
of the two fluxes (note that we actually assume the absolute values of the fluxes):
\be
\langle j\rangle = j_z^{e} + \langle j_z^{\gamma}\rangle = {v + \cos\alpha_k \over V}\,,\label{j-serbo}
\ee
where $\tan\alpha_k = \kappa/k_z$.
We think that a more appropriate definition of flux
should take into account not only the $z$-components of the individual fluxes but also
the relative lateral motion of the two waves. Indeed, the sole purpose of calculating the flux factor
is, classically speaking, to derive the volume swept by one particle in the ``gas'' 
of opposing particles and find how many ``attempts'' at collision are made per unit time. 
Therefore, we propose the following general definition of the integrated flux factor $I_{tw}$
for the single-twisted case:
\be 
I_{tw} = \int {d\phi_k \over 2\pi} I_{PW}(\bk,\bp)\,. \label{flux-1tw}
\ee
Note that it is well defined for any transverse momentum $\bp$ of the opposing plane wave.
In the specific case considered above this definition gives 
$\langle j\rangle = (1 + v\cos\alpha_k)/V$, which differs from (\ref{j-serbo}).
This definition can be also generalized to the double-twisted case:
\be 
I_{2tw} = \int {d\phi_k \over 2\pi}\, {d\phi_p \over 2\pi} I_{PW}(\bk,\bp)\,. \label{flux-2tw}
\ee

\section{Single-twisted cross section}\label{section-single}

We start with a usual $2\to n$ collision in which both incoming particles
are described by plane waves with definite four-momenta $k$ and $p$.
The final system $X$ is also treated as a collection of plane waves
with the total momentum $p_X$.
The invariant amplitude of this process is denoted by ${\cal M}(\bk,\bp)$, 
where the transverse momenta of the initial particles are indicated explicitly. 

The cross section of this process is calculated according to the standard rules.
When squaring the scattering matrix element 
\be
S_{PW} = i(2\pi)^4\delta^{(4)}(k+p-p_X)\cdot {\cal M}(\bk,\bp)\,,\label{S-PW}
\ee
we re-interpret the square of the delta-function as
\be
\left[\delta^{(4)}(k+p-p_X)\right]^2 = \delta^{(4)}(k+p-p_X)\cdot {\pi R^2 L_z T \over (2\pi)^4}\,.
\ee
Using the plane-wave normalization factors (\ref{norm-PW}) for all the initial and final particles,
we get the cross section
\be
d\sigma_{PW}(\bk,\bp) = {(2\pi)^4\delta(E_i-E_f)\delta(p_{zi}-p_{zf})\delta^{(2)}(\bk+\bp-\bp_X) 
\over 4 I_{PW}}|{\cal M}(\bk,\bp)|^2\cdot d\Gamma_X\,.
\label{xsectionPW}
\ee
For future convenience the delta-function is explicitly broken into the longitudinal and transverse parts.
As usual, we can extract from the final phase space integration measure $d\Gamma_X$ 
the integral over the total transverse momentum $\bp_X$, $d\Gamma \equiv d^2\bp_X d\Gamma'_X$,
and write the cross section as
\be
d\sigma_{PW}(\bk,\bp) = {(2\pi)^4\delta(E_i-E_f)\delta(p_{zi}-p_{zf})
\over 4 I_{PW}}|{\cal M}(\bk,\bp)|^2\cdot d\Gamma'_X\,.
\label{xsectionPW2}
\ee

Now we recalculate the cross section for the case when the first particle is in the twisted state $|\kappa,m\rangle$.
We apply prescription (\ref{twisted-def}) to the $S$-matrix element:
\be
S_{tw} = \int {d^2 \bk \over (2\pi)^2} a_{\kappa m}(\bk) S_{PW}(\bk,\bp)\,,\label{passage}
\ee
as originally suggested in \cite{serbo1}.
The plane-wave $S$-matrix (\ref{S-PW}) contains a delta-function of transverse momenta, 
which makes it possible
to simplify the square of the twisted $S$-matrix element as
\bea
|S_{tw}|^2 &=& \int {d^2 \bk \over (2\pi)^2} {d^2 \bk' \over (2\pi)^2} a_{\kappa m}(\bk) a^*_{\kappa m}(\bk') 
S_{PW}(\bk,\bp)S^*_{PW}(\bk',\bp)\nonumber\\
&\propto &  \int {d^2 \bk \over (2\pi)^2} {d^2 \bk' \over (2\pi)^2} a_{\kappa m}(\bk) a^*_{\kappa m}(\bk') 
\delta^{(2)}(\bk+\bp-\bp_X) \delta^{(2)}(\bk'+\bp-\bp_X) {\cal M}(\bk,\bp) {\cal M}^*(\bk',\bp)\nonumber\\
&=& \int {d^2 \bk \over (2\pi)^4} a_{\kappa m}(\bk) a^*_{\kappa m}(\bk) \delta^{(2)}(\bk+\bp-\bp_X) |{\cal M}(\bk,\bp)|^2\,.
\eea
The square of $a_{\kappa m}(\bk)$ contains a radial delta-function squared, which is reinterpreted as
\be
\left[\delta(\kappa-|\bk|)\right]^2 = \delta(\kappa-|\bk|)\cdot \delta(0) \longrightarrow \delta(\kappa-|\bk|) {R\over \pi}\,.
\label{reinterpreted}
\ee
This prescription comes from the observation that at large but finite $R$ and at $\kappa=|\bk|$ 
the radial delta-function is regularized as
\be
\delta(0) = \int_0^\infty rdr [J_{m}(\kappa r)]^2 \to 
\int_0^R rdr [J_{m}(\kappa r)]^2 \approx {R \over \pi}\,.
\ee
see \cite{serbo1,serbo2}.
Therefore, with all the normalization factors (\ref{norm-PW}) and (\ref{norm-tw2}) the 
single-twisted cross section takes form
\be
d\sigma_{tw} = \int {d^2 \bk \over 2\pi} \, {I_{PW}(\bk,\bp) \over I_{tw}}\, {\delta(\kappa-|\bk|) \over \kappa} \cdot d\sigma_{PW}(\bk,\bp)
= \int {d\phi_k \over 2\pi}\, {I_{PW}(\bk,\bp) \over I_{tw}}\, d\sigma_{PW}(\bk,\bp)\,, \label{sigma-tw}
\ee
see Section~\ref{subsection-flux} for the definition of the fluxes. 
Note that in the paraxial approximation one can replace the ratio of the fluxes by the unity.

A couple of remarks concerning this result are in order. In the usual case of a plane wave collision, 
the total initial and, therefore, final momenta are fixed. This is highlighted by the presence of the transverse delta-function
in (\ref{xsectionPW}) and by implicit correlations among transverse momenta of the final particles
inside $d\Gamma'_X$ in (\ref{xsectionPW2}). In the process involving a twisted particle, the total final momentum
is not fixed, making the final particles less correlated.
Although all the plane waves $|PW(\bk)\rangle$ which constitute the initial twisted state $|\kappa,m\rangle$ 
are summed up coherently 
at the amplitude level, the final states with different momenta do not interfere, and this breaks the coherence.
As a result, the twisted particle cross section is expressed via an {\em incoherent} superposition of 
the cross sections induced by each initial plane wave. This is precisely what (\ref{sigma-tw}) displays.

The previous paragraph contains one additional subtlety. 
When saying that the final states with different momenta do not interfere, we implicitly assume
that the final state is detected by a usual detector, which measures the linear momentum but not
the OAM. On the contrary, if the final particles were detected by a hypothetical ``coherent detector''
sensitive to a coherent superposition of final states with distinct momenta,
or if the production of particles in the reaction we consider is followed by another process which is OAM-selective,
then the coherence would be restored. Thus, the coherence is not actually destroyed in the scattering process,
but remains hidden.

The second remark concerns the angular region contributing to the integral (\ref{sigma-tw}).
In the fully differential case, that is when we fix the momenta of all the final particles,
the transverse delta-function inside $d\sigma_{PW}$ assures that there is only one value of $\phi$ 
that contributes to the integral. At this level of consideration, representing the cross section
as an angular integral might look somewhat misleading.
However if this delta-function is killed by the integration over all allowed $\bp_X$, 
or equivalently by integrating over one of the final particles, then the integral receives 
contributions from all angles $\phi_k$.
Note that the angular integral representation is also justified even in the fully differential case
if the detector resolution of the final particles' momenta is worse than $\kappa$.

Let us also show a slightly different derivation of (\ref{sigma-tw}).
We first explicitly perform the integration in (\ref{passage}), which is effectively killed by the 
transverse part of the delta-function, keeping the scattering amplitude essentially unchanged:
\be
\int {d^2 \bk \over (2\pi)^2} a_{\kappa m}(\bk) \delta^{(2)}(\bk+\bp-\bp_X)\cdot {\cal M}(\bk,\bp)
= {(-i)^m \over(2\pi)^{3/2}} e^{im\phi_q}{ \delta(\kappa-q) \over \sqrt{\kappa}}\cdot {\cal M}(\bq,\bp)\,,\label{transverseTW}
\ee
where $\bq \equiv \bp_X - \bp$.
Note that after the integration the transverse momentum $\bk$ cannot appear in the amplitude any more.
However, since the other particles are plane waves with well defined momenta, the vector $\bq$
with modulus $q$ and azimuthal angle $\phi_q$ 
is also well defined and can be used instead of $\bk$ everywhere.
Squaring (\ref{transverseTW}), we again encounter the square of the radial delta-function which is reinterpreted as
in (\ref{reinterpreted}). 
The cross section then becomes
\be
d\sigma_{tw} = {(2\pi)^4\delta(E_i-E_f)\delta(p_{zi}-p_{zf}) |{\cal M}(\bq,\bp)|^2
\over 4 I_{tw}}\cdot d\Gamma_X \cdot {\delta(\kappa-q) \over 2\pi \kappa}\,.\label{xsectionTW}
\ee
The integral over the overall transverse momentum present in $d\Gamma_X$ kills the radial delta-function:
\be
\int d^2\bp_X {\delta(\kappa-q) \over 2\pi \kappa} = 
\int d^2\bq {\delta(\kappa-q) \over 2\pi \kappa} = \int {d\phi_q \over 2\pi}\,,
\ee
and one recovers the result (\ref{sigma-tw}).

One can make several observations concerning (\ref{sigma-tw}):
\begin{itemize}
\item
The cross section is $m$-independent. 
\item
The initial twisted particles effectively perform the angular averaging of the plane wave cross-section.
\item
There is no smallness associated with non-zero $m$. 
\item
There is no small factor associated with small $\kappa$.
\end{itemize}

Let us now consider the case when the twisted particle is not in an eigenstate
of the OAM operator, but in a superposition
of states $|\kappa,m\rangle$ with equal $\kappa$ but different $m$.
For example, consider the twisted state of the form $a|\kappa,m\rangle + a'|\kappa,m'\rangle$,
with $|a|^2+|a'|^2=1$.  Repeating the same calculation we encounter an interference term in the cross section:
\be
d\sigma^{\Delta m}_{tw} = 
\int {d\phi_k \over 2\pi}\, {I_{PW}(\bk,\bp) \over I_{tw}} \cos(\Delta m\,\phi_k + \alpha)\,d\sigma_{PW}(\bk,\bp)\,,\label{xsectionTWint}
\ee
where $\Delta m = m-m'$ and $\alpha$ is the relative phase between the two complex coefficients $a$ and $a'$.
The cross section for such an initial state takes the following form:
\be 
d\sigma = d\sigma_{tw} + 2|aa'|\, d\sigma^{\Delta m}_{tw}\,.
\ee

Results (\ref{sigma-tw}) and (\ref{xsectionTWint}) mean that if the initial particle
can be prepared in a twisted state with an adjustable superposition of different $m$, 
a Fourier analysis of the cross section with respect to the initial azimuthal angle can be performed.
In principle, the same analysis can be done with plane waves, but it would require
making several experiments with different angles of the initial particle $\bk$ and then extracting the Fourier components
via the partial wave analysis. From the experimental view, it is likely that systematics 
of the two schemes can be different, which makes them complementary to each other.

\section{Double-twisted cross section}\label{section-double}

\begin{figure}[!htb]
   \centering
\includegraphics[width=12cm]{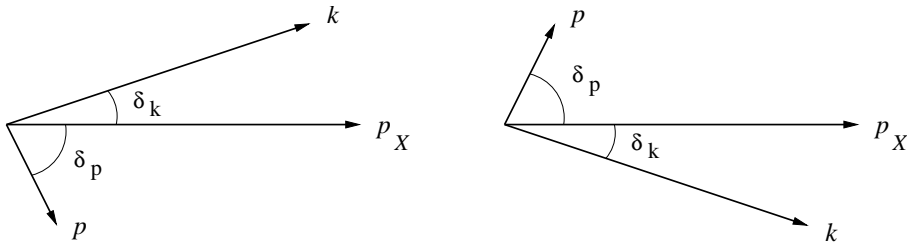}
\caption{Two kinematical configurations of the transverse momenta $\bk$ and $\bp$ of fixed absolute values
that sum up to the vector $\bp_X$.}
   \label{fig1}
\end{figure}

Let us now consider collision of two initial twisted particles. 
In this work we do not aim at a systematic study of this case,
but rather outline some new features can be expected in such circumstances.
Therefore we consider the simplest set-up, in which the two colliding particles are described
by twisted states $|\kappa,m\rangle$ and $|\eta,n\rangle$ defined with respect
to the same quantization axis $z$:
\be
|\kappa,m\rangle = \int {d^2\bk \over(2\pi)^2}a_{\kappa m}(\bk) |PW_1(\bk)\rangle \,,\quad
|\eta,n\rangle = \int {d^2\bp \over(2\pi)^2}a_{\eta n}(\bp) |PW_2(\bp)\rangle \,,
\ee
with the same functional form of the projectors $a$ as before.
Here, the subscripts $1$ and $2$ refer to the first and second colliding particles.
The ``double-twisted'' version of (\ref{passage}) is 
\be
S_{2tw} = \int {d^2 \bk \over (2\pi)^2}\, {d^2 \bp \over (2\pi)^2} a_{\kappa m}(\bk) 
a_{\eta n}(\bp) S_{PW}(\bk,\bp)\,,\label{passage2}
\ee
and its square is proportional to
\bea
&&\int {d^2 \bk\,d^2 \bp\, d^2 \bk'\,d^2 \bp' \over (2\pi)^8} a_{\kappa m}(\bk) a_{\eta n}(\bp)
a^*_{\kappa m}(\bk') a^*_{\eta n}(\bp')\nonumber\\
&& \times \ \delta^{(2)}(\bk+\bp-\bp_X) \delta^{(2)}(\bk'+\bp'-\bp_X) {\cal M}(\bk,\bp) {\cal M}^*(\bk',\bp')\,.
\label{2twistedS}
\eea
Let us consider the kinematical restrictions imposed by the delta-functions entering this expression
and compare them with the result $\bk=\bk'$ found in the previous Section.
Here, too, the moduli of the momenta are fixed and pairwise equal: $|\bk|=\kappa=|\bk'|$ and $|\bp|=\eta=|\bp'|$.
Besides, the two pairs of momenta sum up to a well-defined $\bp_X$.
For each pair there are {\em two} possibilities satisfying these conditions, shown in Fig.~\ref{fig1}, which are
mirror reflections of each other with respect to the direction of $\bp_X$.
Therefore, the integral (\ref{2twistedS}) receives contributions from two kinematical configurations:
\bea
\mbox{direct:}&& \bk'=\bk\,,\ \bp'=\bp\,,\nonumber\\
\mbox{reflected:}&& \bk'=\bk^* \equiv - \bk + 2(\bk\bn_X)\bn_X \,,\ \bp'=\bp^* \equiv - \bp + 2(\bp\bn_X)\bn_X\,,\label{2configurations}
\eea
with $\bn_X \equiv \bp_X/|\bp_X|$.
As we will see below, in contrast to the ``single-twisted'' case,
here the existence of two configuration for any given $\bp_X \not = 0$ 
leads to a residual coherence between different plane wave components in the twisted state.

Our goal now is to see whether the double-twisted cross section can be expressed 
as an angle-averaged plane wave cross section similarly to (\ref{sigma-tw}).
To this end let us first study a generic expression
\be
J=\int d\phi_k\, d\phi_p\,  \delta^{(2)}(\bk+\bp-\bp_X)\cdot f(\bk,\bp)\,,\label{integralJJ}
\ee
where $|\bk|=\kappa$ and $|\bp|=\eta$ are fixed and 
$f(\bk,\bp)$ is a continuous function of the incoming particles' momenta.
Due to the delta-function, the integral $J$ receives contributions only from two points, shown in
Fig.~\ref{fig1}. At these points the azimuthal angles $\phi_k$ and $\phi_p$ take specific values:
\be
\phi_k = \phi_X \pm \delta_k\,,\quad \phi_p = \phi_X \mp \delta_p\,,
\ee
where $\phi_X$ is the azimuthal angle of $\bp_X$ and
\be
\delta_k = \arccos\left({\bp_X^2 + \kappa^2 -\eta^2 \over 2|\bp_X|\kappa}\right)\,,\quad 
\delta_p = \arccos\left({\bp_X^2 - \kappa^2 +\eta^2 \over 2|\bp_X|\eta}\right)
\ee
are functions of the absolute values of the momenta.
Let us denote the values of $f(\bk,\bp)$ at these two points as $f_+$ and $f_-$, respectively.
In Appendix~\ref{appendixA} we derive the following result
\be
J = {f_+ + f_- \over 2\Delta} = \int d\phi_k\, d\phi_p \delta^{(2)}(\bk+\bp-\bp_X)\cdot {f_+ + f_- \over 2}\,,\label{Jresultagain}
\ee
where $\Delta$ is the area of the triangle of sides $|\bp_X|$, $\kappa$, $\eta$.
It allows us to rewrite the square of the integral $J$ as
\bea
|J|^2 &=& \int d\phi_k\, d\phi_p\, \delta^{(2)}(\bk+\bp-\bp_X) f(\bk,\bp) {f^*_+ + f^*_- \over 2\Delta}\nonumber\\
&=& {1 \over 4\Delta} \int d\phi_k\, d\phi_p\, \delta^{(2)}(\bk+\bp-\bp_X)
\left[|f_+|^2+|f_-|^2+2\mathrm{Re}(f_+f_-^*)\right]\nonumber\\
&=& {1 \over 2\Delta} \int d\phi_k\, d\phi_p\, \delta^{(2)}(\bk+\bp-\bp_X)
\left[|f(\bk,\bp)|^2 + \mathrm{Re}f(\bk,\bp)f^*(\bk^*,\bp^*)\right]\,,
\eea
where $\bk^*$ and $\bp^*$ are given by (\ref{2configurations}).

These results can be directly applied to the integral (\ref{2twistedS}) if we note that it has the form of $|J|^2$ with the function
$$
f(\bk,\bp) = {1 \over (2\pi)^3 \sqrt{\kappa\eta}} e^{im\phi_k+in\phi_p} {\cal M}(\bk,\bp)\,.
$$
The integral (\ref{2twistedS}) then takes form
\bea
&&{1 \over (2\pi)^6 \sin(\delta_k+\delta_p)} 
\int d\phi_k d\phi_p \delta^{(2)}(\bk+\bp-\bp_X) 
\nonumber\\ 
&&
\qquad\times \left\{|{\cal M}(\bk,\bp)|^2 +\mathrm{Re}\left[e^{2im(\phi_k-\phi_X) + 2in(\phi_p-\phi_X)}{\cal M}(\bk,\bp){\cal M}^*(\bk^*,\bp^*)\right]\right\}\,.
\eea
Bringing together all the normalization coefficients, we finally arrive at the following representation for 
the double-twisted cross section:
\be
\label{2tw-3}
d\sigma_{2tw} = {1 \over 8\pi \sin(\delta_k+\delta_p)} 
\int d\phi_k\, d\phi_p\, {I_{PW}(\bk,\bp) \over I_{2tw}}\, \left[d\sigma_{PW}(\bk,\bp) + d\sigma'(\bk,\bp)\right]\,,
\ee
where the flux $I_{2tw}$ is given by (\ref{flux-2tw}) and 
\be
d\sigma'(\bk,\bp) = {(2\pi)^4\delta(E_i-E_f)\delta(p_{zi}-p_{zf})
\over 4 I_{PW}} \mathrm{Re}\left[e^{2im(\phi_k-\phi_X) + 2in(\phi_p-\phi_X)} 
{\cal M}(\bk,\bp){\cal M}^*(\bk^*,\bp^*)\right]\cdot d\Gamma'_X\,.
\label{novelsigma}
\ee
We see that the ``direct'' contribution yields the standard cross section, 
while the ``reflected'' contribution gives rise to the novel quantity $\sigma'(\bk,\bp)$.
This quantity describes the {\em auto-correlation} of the amplitude
and is absent in the plane wave case.

Let us also see how (\ref{2tw-3}) simplifies in the case when
${\cal M}(\bk,\bp)={\cal M}(\bk^*,\bp^*)$. The novel quantity reduces to the usual cross section,
$d\sigma'=\cos[2(m\delta_k-n\delta_p)]d\sigma_{PW}$, and we obtain
\be
\label{2tw-4}
d\sigma_{2tw} = {\cos^2(m\delta_k-n\delta_p) \over 4\pi\sin(\delta_k+\delta_p)}  
\int d\phi_k\, d\phi_p\, {I_{PW}(\bk,\bp) \over I_{2tw}}\, d\sigma_{PW}(\bk,\bp)\,.
\ee
It is only in this case that the double-twisted cross section is expressed via
the angular integral of the plane wave cross section.

As in the single-twisted case, we note that if all the final momenta
were fixed and if the detector had an infinitely good momentum resolution,
there would be just two points contributing to the angular integral (\ref{2tw-3}).
If at least one of these conditions is broken, the integrand extends
over the full integration range.

Similarly to the single-twisted case, the double-twisted cross section
stays finite even when the conical momentum spreads $\kappa$ and $\eta$
are very small. However in contrast to the single-twisted case, the cross section now depends also 
on the orbital angular momenta $m$ and $n$. This dependence comes solely from the
correlation term in (\ref{2tw-3}). If the cross section is 
averaged with a sufficiently smooth function of the initial or final momenta,
then in the limit of large $m$ and $n$ the correlation term gets suppressed
by rapid oscillations, and the double-twisted cross section
stays approximately $m$ and $n$ independent. 

Certainly, one can also consider double-twisted cross sections
with twisted states in superposition
of different $m$ and $n$. We leave a detailed study of this situation
for a future work.

Finally we reiterate the point that for this particular version of double-twisted
scattering we chose the simplest possible set-up, in which
both twisted states are defined with respect to the same $z$ axis.
One can also study what would happen if two different axes,
either parallel or not, were used. 
Answering this question requires a more elaborate formalism.

\section{Two-particle decay of a twisted scalar}\label{section-decay}

\subsection{Preliminary remarks}\label{section-problem}

Previous two Sections were devoted to the cases when the twisted particles
appeared only in the initial state, while the final state $X$ was assumed to be describable with the plane waves.
Let us now consider the case when at least one of the final particles is also twisted.
This is exactly the situation that arises in the original suggestion of \cite{serbo1,serbo2}
to use Compton backscattering of the twisted optical photons to produce final high-energy OAM photons.
The key question here is to what extent the twisted parameters of the initial state $|\kappa,m\rangle$ 
determine the parameters of the final twisted state $|\kappa',m'\rangle$.
Here again both twisted states are defined with respect to the same $z$-axis.
In the case of strictly backward scattering the conservation of the orbital momentum
ensures that $\kappa'=\kappa$, $m'=m$, \cite{serbo1}.
In this Section we study how this result changes for the non-forward scattering.
Note that we use the term ``forward'' for any process with zero transverse momentum transfer, 
i.e. both for strictly forward and strictly backward kinematics.

In order to focus on the final state kinematics and to avoid possible complications coming from a non-trivial matrix element,
we would like to answer the above question in the context of the simplest possible problem:
the decay of a twisted scalar particle with mass $M$ into a pair of massless distinguishable particles
due to the cubic interaction $g\cdot \Phi\phi_1\phi_2$.
The momenta of the initial and final particles are $p$, $k_1$ and $k_2$, respectively.
We will calculate the decay width in the center of mass frame defined by $p_z=0$. 
This is not the true rest frame because due to the transverse motion 
a twisted particle is never at rest. 
To make the presentation more pedagogical, we will first calculate the decay rate 
when both particles in the final state are plane waves, 
then for the plane wave plus twisted final state, and finally
for the case when both final particles are twisted.
Although the total decay width must be the same in all these cases, 
the differential decay rates will be rather different.

\subsection{Two plane waves}

Again, let us first recall the standard calculation for the case
when all the particles including the initial one are plane waves.
The $S$-matrix is given as usual by
$S = i(2\pi)^4\delta^{(4)}(p-k_1-k_2)\cdot g$.
With the plane wave normalization coefficients for all the particles,
the differential decay rate for a particle at rest is
\bea
d\Gamma &=& {(2\pi)^4 g^2 \delta^{(4)}(p-k_1-k_2) VT \over T}\cdot (N_{PW}^2)^3
\cdot dn_{PW}(k_1) dn_{PW}(k_2) \nonumber\\
&=& {g^2\over (2\pi)^2} {\delta(M - \omega_1-\omega_2) \over 8M\omega_1\omega_2} d^3k_1\,,\label{PWdecaydiff}
\eea
so that the total width is
\be
\Gamma = {g^2 \over 16\pi M}\,.\label{PWdecay}
\ee

Now we repeat this calculation for the initial twisted state $|\kappa,m\rangle$, 
while keeping the plane wave basis the final particles.
The $S$-matrix is 
\bea
S &=& i(2\pi)^4 g\,\delta(E-\omega_1-\omega_2)\delta(k_{1z}+k_{2z}) 
\int {d^2 \bk \over (2\pi)^2} a_{\kappa m}(\bk) \delta^{(2)}(\bk-\bk_1-\bk_2)
\label{Stwisted1}\\
&=& i(2\pi)^4 g\, \delta(E-\omega_1-\omega_2)\delta(k_{1z}+k_{2z})
{(-i)^m \over(2\pi)^{3/2}} e^{im\phi_{12}}{ \delta(\kappa-k_{12}) \over \sqrt{\kappa}} \,,
\eea
where $k_{12} \equiv \sqrt{\bk_1^2 + \bk_2^2 + 2|\bk_1| |\bk_2|\cos(\phi_1-\phi_2)}$
and $\phi_{12}$ is the angle of the 2D vector $\bk_{12}$ w.r.t. some axis $x$. 
Similaly to the single scattering cross section, 
this phase factor is inessential and $m$ disappears in the decay rate.

The square of $\delta(\kappa-k_{12})$ is treated as in (\ref{reinterpreted}),
and with the appropriate normalization factors taken into account, the decay rate has the form
\bea
d\Gamma &=& {(2\pi)^3 g^2 \over 8 E \omega_1\omega_2 \cdot T}\delta(E-\omega_1-\omega_2)\delta(k_{1z}+k_{2z})T L_z 
{\delta(\kappa-k_{12})\over \kappa}{R\over \pi} \cdot {1\over V^2} {\pi \over RL_z}\cdot 
{Vd^3k_1 \over (2\pi)^3} {Vd^3k_2 \over (2\pi)^3}\nonumber\\
&=& {g^2 \over (2\pi)^3 } {\delta(\kappa-k_{12}) \over \kappa}{\delta(E-\omega_1-\omega_2) \over 8 E \omega_1\omega_2 }\, dk_{z}
\,d^2\bk_1\, d^2\bk_2\,.
\eea
For the transverse integral we write
\be
\int d\phi_2 \,{\delta(\kappa-k_{12}) \over \kappa} = 
2 \int d\phi_2 \,\delta\left[\kappa^2 - \bk_1^2-\bk_2^2-2|\bk_1||\bk_2|\cos(\phi_1-\phi_2)\right] = {1 \over \Delta}\,,
\label{angular-integral}
\ee
where $\Delta$ is the area of the triangle with sides $\kappa$, $|\bk_1|$ and $|\bk_2|$, see Appendix~\ref{appendixA}.
As usual, the energy delta function can be killed by the $k_{z}$ integration
\bea
\int dk_{z} {\delta(E-\omega_1-\omega_2) \over \omega_1\omega_2} = {2 \over E k_z^*}\,,
\eea
where 
\be
k_z^* = {1 \over 2E}\sqrt{E^4+\bk_1^4+\bk_2^4 - 2(E^2\bk_1^2 + E^2 \bk_2^2 + \bk_1^2 \bk_2^2)}\,.
\ee
The decay rate becomes
\be
d\Gamma = {g^2 \over 4\pi^2} {1 \over 4 E^2 k_z^*} {|\bk_1| d|\bk_1|\, |\bk_2| d|\bk_2|\over \Delta} \,.\label{dGamma-PWPW}
\ee
The integration region over $|\bk_1|$ and $|\bk_2|$ is defined by the requirement that, in addition to (\ref{triangle-rule}), 
the longitudinal momentum $k_z^*$ is well-defined, which cuts the rectangular shape shown in Fig.~\ref{fig-region} in Appendix~\ref{appendixA}.
It is conveniently described with variables 
\be
x = {|\bk_1|-|\bk_2| \over \kappa}\,,\quad z = {|\bk_1|+|\bk_2| \over \kappa}\,,
\quad x\in [-1,1]\,,\quad z\in [1,z_{max}]\,,\quad z_{max} \equiv {E\over \kappa} > 1\,.\label{limits-xz}
\ee 
In these variables 
\be
\Delta = {\kappa^2 \over 4}\sqrt{(z^2-1)(1-x^2)}\,,\quad
k_z^* = {E\over 2}\sqrt{\left(1-{x^2 \over z_{max}^2}\right)\left(1-{z^2 \over z_{max}^2}\right)}\,,
\ee
and the decay rate takes form
\be
d\Gamma = {g^2 \over 16 \pi^2 E}\cdot {(z^2-x^2)dzdx \over \left[(z_{max}^2-z^2)(z_{max}^2-x^2)(z^2-1)(1-x^2)\right]^{1/2}}\,.
\ee
This integral can be taken exactly, 
and it gives the decay width of the twisted scalar particle
\be
\Gamma = {g^2 \over 16\pi E}\,,\label{twisted-decay}
\ee
which is a very natural result. In the limit $\kappa \to 0$, we recover the plane wave 
decay width (\ref{PWdecay}). 

Of course, one could also arrive at (\ref{twisted-decay}) 
just by using our previous results concerning the single-twisted cross section modified to the case of
one initial particle. However the detailed derivation given here is needed to understand
what changes when the final state includes twisted particles.

\subsection{Twisted state plus plane wave}\label{subsection-twpw}

Let us now describe the final state as twisted state plus a plane wave:
\be
|\kappa,m\rangle \to |\kappa_1,m_1\rangle + |PW(\bk_2)\rangle\,.
\ee
Since the full decay width cannot depend on the basis we choose for the final particles,
we must recover the same result (\ref{twisted-decay}) in this basis.
In addition to that, we also want to know how the final twisted state parameters $\kappa_1$ and $m_1$
are related to the initial state parameters $\kappa$ and $m$. 

The $S$-matrix is now 
\be
S=\int {d^2\bk \over(2\pi)^2} {d^2\bk_1 \over(2\pi)^2}
a^*_{\kappa_1 m_1}(\bk_1) a_{\kappa m}(\bk)\, S_{PW} \,.\label{S_tw} 
\ee
It can be written as 
\be
S = i (2\pi)^4 g \delta(E-\omega_1-\omega_2)\delta(k_{z1}+k_{z2}) 
\cdot {\cal I}_{m,m_1}(\kappa,\kappa_1,\bk_2)\,.
\ee
where the master integral ${\cal I}_{m,m_1}(\kappa,\kappa_1,\bk_2)$ is defined as
\be
{\cal I}_{m,m_1}(\kappa,\kappa_1,\bk_2) = \int {d^2\bk \over(2\pi)^2} {d^2\bk_1 \over(2\pi)^2}
a^*_{\kappa_1 m_1}(\bk_1) a_{\kappa m}(\bk) \delta^{(2)}(\bk - \bk_1 -\bk_2)\,.\label{master}
\ee

Let us first calculate this integral in the strictly forward case $\bk_2=0$:
\bea
{\cal I}_{m,m_1}(\kappa,\kappa_1,0) &=& 
{i^{m_1-m} \over(2\pi)^3} \int d^2\bk\, d^2\bk_1 e^{im\phi-im_1\phi_1}{\delta(|\bk|-\kappa) \over\sqrt{\kappa}}
{\delta(|\bk_1|-\kappa_1) \over\sqrt{\kappa_1}} \delta^{(2)}(\bk-\bk_1)\nonumber\\
&=& {i^{m_1-m} \over(2\pi)^3}\sqrt{\kappa\kappa_1}\int d\phi\, d\phi_1\, e^{im\phi - im_1\phi_1}\, \cdot
2\delta(\bk^2-\bk_1^2)\delta(\phi-\phi_1)\nonumber\\
&=& {1 \over(2\pi)^2}\delta(\kappa-\kappa_1) \delta_{m,m_1}\,,\label{master-forward}
\eea
which was first obtained in \cite{serbo1}.
This result implies that the twisted state quantum number are transferred from the initial
to the final twisted particle without any change.
The differential decay rate is
\bea
d\Gamma &=& {g^2 \over (2\pi)^2}\, {\delta(E-\omega_1-\omega_2)\delta(k_{1z}+k_{2z}) \over 8E\omega_1\omega_2}
\delta_{m,m_1}\delta(\kappa-\kappa_1)\cdot d\kappa_1 dk_{1z} d^3k_2\nonumber\\
&=& {g^2 \over (2\pi)^2} {\delta(E-\omega_1-\omega_2) \over 8 E \omega_1 \omega_2 } d^3 k_2
= {g^2 \over (2\pi)^2} {d^2\bk_2 \over 4 E^2 k_z^*}\,,\label{result-strictly-forward}
\eea
and we stress that this result is applicable only at $\bk_2 =0$.

In the general non-forward case the master integral (\ref{master}) can be rewritten as
\be
{\cal I}_{m,m_1}(\kappa,\kappa_1,\bk_2) = {i^{m_1-m} \over(2\pi)^3}\sqrt{\kappa\kappa_1}\int d\phi d\phi_1\, 
e^{im\phi - im_1\phi_1}\, \cdot \delta^{(2)}(\bk-\bk_1-\bk_2)\,.\label{master2}
\ee
There are two ways to look at this integral. First, using the results of Appendix~\ref{appendixA} we obtain
\be
{\cal I}_{m,m_1}(\kappa,\kappa_1,\bk_2) = {i^{m_1-m} \over(2\pi)^3}\sqrt{\kappa\kappa_1}
e^{i(m-m_1)\phi_2}{\cos[m_1\delta_1 - (m-m_1)\delta_2] \over \Delta}\,,\label{master4}
\ee
where $\Delta$ is the same as in (\ref{Delta-def}) with $\bk_1^2$ replaced by $\kappa_1^2$.
Additionally, we can also rewrite 
\be
\delta^{(2)}(\bk-\bk_1-\bk_2) = {1\over(2\pi)^2} \int d^2\br\,  
e^{i\br\bk - i \br \bk_1 - i \br\bk_2}
\label{method1start}
\ee
and represent the master integral as an integral over a triple product of Bessel functions which is
useful in certain circumstances:
\be
{\cal I}_{m,m_1}(\kappa,\kappa_1,\bk_2)  = {i^{m_1-m} e^{i(m-m_1)\phi_2}\over(2\pi)^2}\sqrt{\kappa\kappa_1}
\cdot 
\int_0^\infty rdr J_m(\kappa r) J_{m_1}(\kappa_1 r) J_{m-m_1}(|\bk_2| r)\,,\label{master3}
\ee
where $\phi_2$ is the azimuthal angle of $\bk_2$.
Note that although we are considering the case with two twisted particles (one in the initial and one in the final state),
an integral over three Bessel function arises automatically.
Comparison of (\ref{master3}) with (\ref{master4}) gives the result for the integral
of the triple Bessel function product. See also \cite{JJJ} for some mathematics
involved in evaluation of this and similar integrals. 

Let us also check the $\bk_2\to 0$ limit of the result (\ref{master4}).
When $|\bk_2|\ll \kappa$, the distribution over $\kappa_1$ spans from
$\kappa - |\bk_2|$ to $\kappa + |\bk_2|$. The angle $\delta_1 \to 0$,
while $\delta_2$ can still be arbitrary. However, in the $\bk_2\to 0$ limit
only $m-m_1=0$ term survives due to $J_{m-m_1}(|\bk_2| r)$, 
so that the cosine in (\ref{master4}) approaches unity.
The analysis of $\Delta$ shows that 
\bea
\lim_{|\bk_2|\to 0} { \sqrt{\kappa\kappa_1} \over \Delta} = 
2\pi \delta(\kappa-\kappa_1)\,,\label{limitk2}
\eea
so that one indeed recovers the strictly forward result for the master integral
given in (\ref{master-forward}). Note that although we wrote simply $|\bk_2|\to 0$,
we imply a very specific combination of this limit with the limit $R\to 0$, an issue
to be discussed in Section~\ref{section-discussion2}.

We continue the calculation of the decay rate in the non-forward case.
After integration over $k_z$, the decay rate can be written as
\be
d\Gamma = 4\pi^3 g^2 {1 \over 4 E^2 k_z^*} \cdot 
|{\cal I}_{m,m_1}(\kappa,\kappa_1,\bk_2)|^2 \cdot {d\kappa_1 \over R}\, d^2\bk_2\,.\label{dGamma_twPW0}
\ee
Note that this decay rate is differential not only in $\kappa_1$ and $\bk_2$ but also in the discrete variable $m_1$;
the full decay width includes integrals over momenta and a summation over all possible $m_1$'s.

A close inspection shows that the immediate integration over $\kappa_1$ or $\bk_2$ 
cannot be done due to singularities of $|{\cal I}|^2$ along the boundaries of the kinematically allowed region.
In contrast to the plane waves in the final state, the denominator now contains $\Delta^2$ instead of just $\Delta$.
Therefore, in terms of variables $x$ and $z$ one encounters singularities of the form
$$
\int_{-1}^1 {dx \over 1-x^2} \quad \mbox{and} \quad \int_1^{z_{max}} {dz \over z^2-1}\,.
$$
Clearly, this is an artefact of the infinite radial integration range in (\ref{master3}). 
If instead we take $R$ to be large but finite,
we expect that a trick similar to (\ref{reinterpreted}) should be at work, 
namely that after regularization $|{\cal I}|^2$ 
would yield $R$ times a less singular function.

This trick does not seem to work for each $m_1$ separately.
However, as we prove in Appendix~\ref{appendixB}, it works for $|{\cal I}|^2$ summed over all
possible $m_1$. In the limit $R\to \infty$ we obtain
\be
\sum_{m_1=-\infty}^{+\infty} |{\cal I}_{m,m_1}(\kappa,\kappa_1,\bk_2)|^2
= {1 \over (2\pi)^5} {R\kappa_1 \over \pi} {1 \over \Delta}\,,
\ee
with the same $\Delta$ as before. The regularization parameter $R$ then disappears
from the result, and the decay rate reads
\be
d\Gamma = {g^2 \over 4\pi^2} {1 \over 4 E^2 k_z^*} \cdot 
{\kappa_1 d\kappa_1\, |\bk_2| d |\bk_2|\over \Delta}\,.\label{dGamma_twPW}
\ee
Comparing (\ref{dGamma_twPW}) with the previous results (\ref{result-strictly-forward}) 
and (\ref{dGamma-PWPW}) leads us to two conclusions.
\begin{itemize}
\item
The transition from the strictly forward to the non-forward cross section/decay rate
consists in replacement
\be
{\delta(\kappa-\kappa_1) \over \kappa_1} \to {1 \over 2\pi\Delta}\,.\label{the_rule}
\ee
\item
The result (\ref{dGamma_twPW}) for a twisted particle in the final state
coincides with the result (\ref{dGamma-PWPW}) for the case when both final particles
are plane waves. 
\end{itemize}
Although these conclusions were drawn for the specific process we consider,
it represents a universal kinematical feature characteristic to all processes involving non-forward scattering 
of twisted states.

\subsection{Two twisted states}

For completeness, let us also recalculate the decay rate in the basis when
both final particles are described by twisted states: $|\kappa_1,m_1\rangle$ and $|\kappa_2,m_2\rangle$.
We remind that all twisted states are defined with respect to the same common $z$ axis.
It turns out that this calculation closely follows the case of 
twisted state plus plane wave just considered. This is not surprising because
the appearance of the triple Bessel integral highlights the fact that when two particles
(one in the initial and one in the final state) are twisted, the third one is automatically
projected from the plane wave onto an appropriately defined twisted state as well.

The $S$-matrix takes the form 
\be
S = i g (2\pi)^{3/2}\delta(E-\omega_1-\omega_2)\delta(k_{z1}+k_{z2}) \cdot \delta_{m,m_1+m_2}
\sqrt{\kappa\kappa_1\kappa_2} \int rdr J_m(\kappa r)J_{m_1}(\kappa_1 r) J_{m_2}(\kappa_2 r)\,,
\ee
and we again encounter the triple Bessel-function integral. 
The decay rate is written as
\be
d\Gamma = {g^2 \over 4\pi^2} {1 \over 4 E^2 k_z^*} \cdot 
{\kappa_1 d\kappa_1\, \kappa_2 d \kappa_2\over \Delta}\,.\label{dGamma_twtw}
\ee
This expression is identical to (\ref{dGamma_twPW}) up to the obvious replacement $|\bk_2| \to \kappa_2$;
its integration over all conical momenta spreads $\kappa_1$, $\kappa_2$ gives again (\ref{twisted-decay}).

\section{Discussion}\label{section-discussion}

\subsection{Potential uses of the twisted cross sections}

In this paper we represented the cross sections for the single-twisted 
and double-twisted processes via the angular integrals
(\ref{sigma-tw}) and (\ref{2tw-3}).
These are rather unconventional quantities, as they involve averaging
of the plane wave cross sections over the azimuthal angles of the {\em initial}
particles at fixed final momenta. 
A further study is needed to see if they can be related to the 
more conventional cross sections, in which the initial momenta are fixed but 
final state azimuthal angles are integrated out.

Using twisted states in superpositions of different orbital angular momenta,
one can perform, in principle, a Fourier-analysis of the differential cross section
as the function of the initial azimuthal angles at fixed final momenta, 
see (\ref{xsectionTWint}). This might be a useful complementary tool for 
the study of various azimuthal asymmetries.

Perhaps, the most intriguing feature we have found arises in the double-twisted 
cross section. We showed that this cross section is sensitive 
not only to the plane wave cross section,
but also to a quantity describing the {\em autocorrelation} 
of the plane wave amplitude.
Such quantity is absent in the conventional plane wave scattering.

Note that it is not the first time the amplitude autocorrelation appears 
in particle physics. In \cite{serbo1992}, where finite beam size effects were analysed,
the product of amplitudes at different momentum space points was also involved.
The essential novelty of the present case is the extra angular factors in 
(\ref{novelsigma}), which offer a more detailed probe of the amplitude.

Finally, let us stress once again that experimental study of these features does not
require large conical momentum spread $\kappa$ and is feasible
even for $\kappa$ in the eV range, which was already realized for energetic electrons
in \cite{twisted-electron,twisted-electron2} and which was proposed in \cite{serbo1} 
for high-energy twisted photons. It is the non-trivial angular dependence that is at work.
However we stress that in order to probe
the autocorrelation function of the amplitude with the momentum offset $\bq$, 
one would need conical momentum spread $\kappa \sim |\bq|$. 

It is natural to ask whether the new degree of freedom offered by twisted
particles, and in particular photons, can be used to gain more insight into the
structure of hadrons. Indeed, the structure of proton remains one of the hottest topics 
in QCD. One of the questions that might be particularly relevant for the present study 
is the contribution of the quark/gluon orbital momentum to the proton spin,
see e.g. reviews \cite{proton-spin}.
It is, therefore, very interesting to know whether the OAM of the initial
photon can be transferred to partons, allowing us to probe the partonic composition of the
proton in a novel way.

Without going into detail, let us make some preliminary observations 
that can be inferred from the results of the present work.

Let us first note that the definition of a twisted state (\ref{twisted-def})
involves an angular functional: 
it picks up a plane wave quantity as a function of the initial azimuthal angle
$\phi$, weights it with $e^{im\phi}$ and integrates
over all angles. This might give an impression that 
the twisted state is a perfect Fourier analyser at level of amplitude.
This impression is wrong because this functional acts not on the amplitudes
itself, but on the $S$-matrix element, which includes a transverse delta-function.
Eq.~(\ref{transverseTW}) shows that this functional partially kills the delta-function, 
while the amplitude remains unchanged. The orbital angular momentum is
just transferred to the overall motion of the final system instead of exciting
internal degrees of freedom.
This somewhat pessimistic conclusion is supported by the result of this paper 
that the single-twisted cross section is just an azimuthal integral of the plane wave cross section.

The situation might change if both the initial photon and proton were twisted.
As we showed above, in this case a residual coherence between two kinematical configuration survives,
which leads to an additional term in the cross section 
proportional to the autocorrelation function of the amplitude.
This term has a non-trivial azimuthal structure and might
represent a new probe of the proton. This issue certainly deserves a dedicated study.

\subsection{Properties of the final twisted particle in a non-forward scattering}\label{section-discussion2}

Let us also discuss in detail what the results of Section~\ref{subsection-twpw} tell us about the values of $\kappa_1$ and $m_1$. 
The differential decay rate (\ref{dGamma_twPW}) 
shows that at large $|\bk_2|\gg \kappa$ the conical momentum spread 
$\kappa_1$ is limited to the interval from $|\bk_2|-\kappa$ to $|\bk_2|+\kappa$ with an inverse square root
singularity at the endpoints. This singularity is integrable, so that the entire interval more or less homogeneously 
contributes to the integral. Since $|\bk_2|$ can be as high as $E$, 
the total decay width is therefore dominated by large $\kappa_1 \gg \kappa$.
In a more complicated process, the decay rate or the cross section
will include the amplitude squared which can serve as a cut-off function.
For example, in the Compton scattering one expects that $\kappa_1$ up to $\sim m_e$ will
contribute to the cross section.

Although we found a result for the decay rate summed over all $m_1$, 
we can trace the main $m_1$-region from the intermediate formulas. The result
is that essentially all $m_1$ from minus to plus infinity are important for the decay rate,
which is in a strong contrast to the strictly forward result $m=m_1$.

Indeed, the distribution over final $m_1$ can be seen in (\ref{dGamma_twPW0}), where
the master integral ${\cal I}_{m,m_1}$ is given by (\ref{master4}). 
For generic transverse momenta, growth of $m_1$ leads to oscillations of the cosine function
with constant amplitude. Thus, when averaging over the entire $\kappa_1$ interval,
one can approximate cosine squared by $1/2$, and the dependence on $m_1$ drops off.
This result holds for any non-zero transverse momentum transfer $|\bk_2|\to 0$.

Since the forward and non-forward $m_1$-distributions are so dramatically different,
a natural question arises whether there is a continuous transition from 
the non-forward to the forward scattering. The answer to this question involves an accurate treatment
of two limits: $|\bk_2|\to 0$ and $R \to \infty$. Let us keep $R$ large but finite, and set $|\bk_2|\to 0$;
then the transition is smooth.
Looking at the triple-Bessel representation of the master integral (\ref{master3}) with the upper limit
replaced by $R$, one sees that the result will begin to significantly decrease only
when the position of the first node of the last Bessel function $J_{m-m_1}(|\bk_2|r)$
falls outside of the integration range, that is for $|\bk_2| \lsim |m-m_1|/R$, where $m_1 \not = m$.
If $m_1=m$, then at $|\bk_2| \ll 1/R$ the last Bessel function can be approximated by the unity.
Therefore, the limit $|\bk_2|\to 0$ taken in (\ref{limitk2}) implies that
$$
|\bk_2|\to 0\quad \mbox{and}\quad R \to \infty\quad \mbox{provided that}\quad  |\bk_2|R \ll 1\,.
$$
If instead $R\to \infty$ at fixed $|\bk_2|$, then the transition of non-forward to forward
results is discontinuous at $|\bk_2|=0$.

In \cite{serbo1} it is claimed with the specific example of the Compton
cross-section that if the transverse momentum transfer is small compared to $\kappa$ 
(in our notation, finite $|\bk_2| \ll \kappa$ at infinite $R$),
then the $m_1$-dependence has a narrow distribution peaked at $m_1=m$.
Our analysis does not support this claim. 

Looking back at the formalism used, we can conclude that 
our result that all $m_1$ essentially contribute to the decay rate/cross section
just reflects the unfortunate choice of the same common axis $z$ for
all the twisted states appearing in the process. 
It does not give a clue of how twisted the final particles are with respect to their own 
propagation axes defined by their average values of the 3-momentum operator. 
Indeed, even a simple non-forward plane wave when expanded in the basis of twisted states
contains all partial waves, see (\ref{PWlimit}). Nevertheless it carries a zero
orbital angular momentum with respect to its own direction of propagation.
Therefore, it appears that a more physically reasonable
quantity is the {\bf ``orbital helicity''}, projection of orbital angular momentum 
on the axis of motion. The relevant question is then how this
``orbital helicity'', not the OAM with respect to a fixed axis, is transferred
from the initial to the final twisted state.
We postpone this question for future studies.

\subsection{Conclusions}

Orbital angular momentum (OAM) is a new degree of freedom, 
which can be used in high-energy physics to gain more insight into properties
of particles and their interactions.
In this paper, focusing on the scalar case, we studied high-energy collisions in which
initial and/or final particles were described by twisted states, 
i.e. carried a non-zero OAM.
We derived expressions for the cross sections for the single-twisted and 
double-twisted cases as well as for the two-particle decay rate of a twisted scalar, 
and observed a number of remarkable features.

The single-twisted differential cross section is represented via the 
plane wave cross section averaged over the azimuthal angle of one of the incoming particles. 
If the initial twisted particle is prepared in a superposition state with different
orbital quantum numbers, a Fourier analysis of the plane wave cross section can be performed.
The expression we found for the double-twisted cross section is more intriguing,
as it involves not only the plane wave cross section, but also 
the autocorrelation function of the amplitude.
We stress that these features do not rely on large conical momentum spread 
in the twisted states, and their experimental study looks feasible 
even with today's technology.

When analyzing the decay rate of a twisted particle, we established the procedure
that allows one to pass from the plane wave to the twisted particle basis for the final state.
These results can be now used, for example, to investigate the Compton
backscattering of the twisted photons in the non-forward region, which was missing in
the original suggestion \cite{serbo1,serbo2}.

Analyzing parameters of the final twisted state, we discovered that there is no smooth
transition from the non-forward to forward scattering.
We attribute this result to the infinite transverse size of twisted states 
and to the deficiency of our formalism in which all the twisted states are defined with respect to 
the same $z$-axis. 
We expect that the problem will disappear if the orbital angular momentum projection 
is calculated not on the fixed reaction axis but on the direction of the outgoing twisted particle. 
Incorporation of this ``orbital helicity'' into the present formalism yet remains to be done.

\section*{Acknowledgements} 
The author is grateful to I.~Ginzburg and V.~Serbo for many stimulating discussions
and to J.-R.~Cudell for valuable comments.
This work was supported by the Belgian Fund F.R.S.-FNRS via the
contract of Charg\'e de recherches, and in part by grants
RFBR 11-02-00242-a and NSh-3810.2010.2.

\appendix

\section{Restricted angular integrals}\label{appendixA}

Here we derive some properties of the restricted angular integrals which appear in the paper. 
The basic integral has the following form
\be
J_0=\int_0^{2\pi} d\phi_1\, d\phi_2\,  \delta^{(2)}(\bk_1+\bk_2-\bk)\label{integralJ_0}
\ee
and corresponds to a situation when two transverse momenta $\bk_1$ and $\bk_2$ of fixed absolute values
sum up to a fixed momentum $\bk$ with modulus $|\bk|=\kappa$ and azimuthal angle $\phi_k$.

Clearly the integral can be non-zero only if it is possible at all
to form a triangle with sides $\kappa$, $|\bk|_1$, and $|\bk_2|$, that is,
when $|\bk_1|$ and $|\bk_2|$ satisfy the ``triangle rules'':
\be
\kappa \le |\bk_1|+|\bk_2|\,,\quad
|\bk_1| \le \kappa+|\bk_2|\,,\quad
|\bk_2| \le \kappa+|\bk_1|\,.\label{triangle-rule}
\ee
\begin{figure}[!htb]
   \centering
\includegraphics[width=6cm]{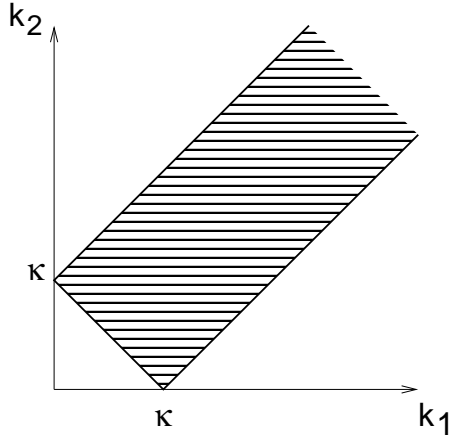}
\caption{The allowed kinematical region of the values of $|\bk_1|$ and $|\bk_2|$ 
for a fixed $\kappa$ defined by the ``triangle rules'' (\ref{triangle-rule}).}
   \label{fig-region}
\end{figure}
This allowed region on the $(|\bk_1|,|\bk_2|)$-plane has the form of a stripe shown in Fig.~\ref{fig-region}.
If all moduli are fixed, the integral (\ref{integralJ_0}) receives contributions only from two points (see Fig.~\ref{fig1} in the main text).
Let us introduce
\be
\delta_1 = \arccos\left({\kappa^2+\bk_1^2 - \bk_2^2 \over 2 \kappa |\bk_1|}\right)\,,
\quad
\delta_2 = \arccos\left({\kappa^2+\bk_2^2 - \bk_1^2 \over 2 \kappa |\bk_2|}\right)\,,\label{deltas}
\ee
Then, the two configurations of transverse momenta correspond to
\be
\phi_k -\phi_1 = \pm \delta_1\,,\quad \phi_k -\phi_2 = \mp \delta_2\,,\label{two-angles}
\ee
so that the signs of $\phi_k -\phi_1$ and $\phi_k -\phi_2$ are always opposite.

To evaluate the integral itself, we rewrite the delta-function as
\be
\delta^{(2)}(\bk_1-(\bk-\bk_2)) = 2 \delta[\bk_1^2 - (\bk - \bk_2)^2]
\,\delta(\phi_1 - \phi_{\bk-\bk_2}) \,.
\ee
The $\phi_1$ integration is then eliminated, and we obtain
\be
J_0 = 2 \int d\phi_2\,\delta\left[\bk_1^2 - \kappa^2-\bk_2 + 2\kappa|\bk_2|\cos(\phi_2-\phi_k)\right] = {1 \over \Delta}\,,
\ee
where
\be
\Delta = {1\over 4}\sqrt{2(\bk_1^2\bk_2^2+\kappa^2 \bk_1^2+\kappa^2 \bk_2^2)
- \kappa^4 - \bk_1^4 - \bk_2^4}\label{Delta-def}
\ee
is the area of the triangle with sides $\kappa$, $|\bk_1|$ and $|\bk_2|$.
When needed, this area can be also rewritten as
\be
\Delta = {1\over 2}|\bk_1| |\bk_2|\sin(\delta_1+\delta_2)\,.
\ee

Let us now consider an integral similar to (\ref{integralJ_0}) but with an extra angular dependence
in the integrand:
\be
J=\int_0^{2\pi} d\phi_1\, d\phi_2\,  \delta^{(2)}(\bk_1+\bk_2-\bk)\cdot f(\phi_1,\phi_2)\label{integralJ}
\ee
with a regular function $ f(\phi_1,\phi_2)$.
Here again the momenta must satisfy the triangle rules (\ref{triangle-rule}) 
and the integral receives contribution only from two points (\ref{two-angles}) in the $(\phi_1,\phi_2)$-space.
Denoting the values of $f(\phi_1,\phi_2)$ at these two points as $f_+$ and $f_-$, respectively,
we obtain
\be
J = {f_+ + f_- \over 2\Delta}\,,\label{Jresult}
\ee
We can also rewrite this result as
\be
J=\int_0^{2\pi} d\phi_1\, d\phi_2\,  \delta^{(2)}(\bk_1+\bk_2-\bk)\cdot {f_+ + f_- \over 2}\,,\label{resultJ2}
\ee
which means that in these circumstances the delta-function
plays the role of a functional that maps a test function $f(\phi_1,\phi_2)$ to its symmetrized
value $(f_+ + f_-)/2$.

In the particular case of $f = \exp(im_1\phi_1 + im_2\phi_2)$, (\ref{Jresult}) take form
\be
J = e^{i(m_1+m_2)\phi_k}{\cos(m_1\delta_1 - m_2\delta_2) \over \Delta}\,,\label{Jresultpartocular}
\ee

\section{Regularization of $|{\cal I}|^2$}\label{appendixB}

Here we calculate the large-$R$ behavior of the 
$m_1$-sum of the squares of the triple-Bessel integral:
\be
\sum_{m_1=-m_{max}}^{+m_{max}} \left[\int_0^R rdr J_m(\kappa r)J_{m_1}(\kappa_1 r) J_{m-m_1}(\kappa_2 r)\right]^2\,,
\label{integral-appendix}
\ee
which appears in the decay rate (\ref{dGamma_twPW0}).
Evaluation of the integral itself with $R\to \infty$ performed in the main text
shows that it can be non-zero only if $\kappa$, $\kappa_1$, $\kappa_2$ satisfy the triangle rules (\ref{triangle-rule}),
i.e. a triangle with these sides can be constructed.
Since $m$ describes the initial state, we take it small and not growing with $R$: 
$m\ll m_{1max} = \kappa_1 R$, while $m_1$ can extend up to $m_{1max}$. 
The final $\kappa_1$, $\kappa_2$ can be much larger than $\kappa$. 

Since the expression (\ref{integral-appendix}) is regularized with large but finite $R$, 
the summation and integration can be interchanged:
\be
\int rdr \, r'dr'\, J_m(\kappa r) J_m(\kappa r')
\sum_{m_1=-m_{1max}}^{+m_{1max}} J_{m_1}(\kappa_1 r) J_{m-m_1}(\kappa_2 r) J_{m_1}(\kappa_1 r') J_{m-m_1}(\kappa_2 r')\,.
\label{sum2}
\ee
Thanks to the properties of the Bessel functions, only $m_1$'s up to $min(\kappa_{1,2}r,\,\kappa_{1,2}r')$
are effectively contributing to this sum; for larger $m_1$ the Bessel functions strongly decrease. 
But $r,r' \le R$, which means that the limits on the summation can in fact be safely extended to the infinity.
Then, the sum of the product of four Bessel functions is treated in the following way:
\bea
&& \sum_{m_1=-\infty}^{+\infty} J_{m_1}(\kappa_1 r) J_{m-m_1}(\kappa_2 r) J_{m_1}(\kappa_1 r') J_{m-m_1}(\kappa_2 r') 
\label{sum4bessels}\\
&& = \sum_{m_1, m'_1=-\infty}^{+\infty} J_{m_1}(\kappa_1 r) J_{m-m_1}(\kappa_2 r) 
J_{m'_1}(\kappa_1 r') J_{m-m'_1}(\kappa_2 r')\cdot \delta_{m_1,m'_1}\nonumber\\
&& = \sum_{m_1, m'_1=-\infty}^{+\infty}  {1 \over 2\pi}\int_0^{2\pi} d\alpha e^{i(m_1-m'_1)\alpha} \,
J_{m_1}(\kappa_1 r) J_{m-m_1}(\kappa_2 r) J_{m'_1}(\kappa_1 r') J_{m-m'_1}(\kappa_2 r')\nonumber\\
&& = {1 \over 2\pi}\int_0^{2\pi} d\alpha 
\left[\sum_{m_1=-\infty}^{+\infty} e^{im_1\alpha} J_{m_1}(\kappa_1 r) J_{m-m_1}(\kappa_2 r)\right]
\left[\sum_{m'_1=-\infty}^{+\infty} e^{-im'_1\alpha} J_{m'_1}(\kappa_1 r') J_{m-m'_1}(\kappa_2 r')\right]\,.\nonumber
\eea
The first sum in the square brackets is calculates as follows:
\bea 
&&\sum_{m_1=-\infty}^{+\infty} e^{im_1\alpha} J_{m_1}(\kappa_1 r) J_{m-m_1}(\kappa_2 r) \nonumber\\
&&= {(-i)^m \over (2\pi)^2}\int d\phi_1\, d\phi_2\, e^{i\kappa_1 r \cos\phi_1 + i \kappa_2 r \cos\phi_2} 
\sum_{m_1=-\infty}^{+\infty}
e^{im_1\alpha + i m_1\phi_1 + i(m-m_1)\phi_2}\nonumber\\
&& =  {(-i)^m \over 2\pi}e^{im\alpha}\int d\phi_1\, e^{im\phi_1}e^{i\kappa_1 r \cos\phi_1 + i k_2 r \cos(\phi_1+\alpha)}\,.\label{sum_ajj}
\eea
The combination of angles and momenta inside the exponential can be expressed as 
\be
\kappa_1\cos\phi_1 + \kappa_2 \cos(\phi_1+\alpha) = |\bk_1+\bk_2|_{\alpha}\cos(\phi_1+\delta\phi)\,,
\ee
where
\be
\quad |\bk_1+\bk_2|_{\alpha} \equiv \sqrt{\kappa_1^2+\kappa_2^2 + 2\kappa_1 \kappa_2\cos\alpha}\,,\quad
\tan\delta\phi = {\kappa_2 \sin\alpha \over \kappa_1 + \kappa_2\cos\alpha}\,.
\ee
Geometrically, $|\bk_1+\bk_2|_{\alpha}$ is the norm of the sum of two vectors 
of moduli $\kappa_1$ and $\kappa_2$ and the relative azimuthal angle $\alpha$.
Therefore, (\ref{sum_ajj}) is
\be
\sum_{m_1=-\infty}^{+\infty} e^{im_1\alpha} J_{m_1}(\kappa_1 r) J_{m-m_1}(\kappa_2 r) = 
e^{im(\alpha-\delta\phi)}\, J_m(|\bk_1+\bk_2|_{\alpha} r)\,.
\ee
This expression can be viewed as a 2D generalization of the well-known addition formula for the Bessel functions
$$
\sum_{m_1=-\infty}^{+\infty} J_{m_1}(x)J_{m-m_1}(y) = J_{m}(x+y)\,.
$$
Now, the second sum differs only by $\alpha\to -\alpha$ (or, alternatively, complex conjugation)
and $r\to r'$. Therefore, the summation (\ref{sum4bessels}) is simplified to
\be
{1 \over 2\pi}\int_0^{2\pi} d\alpha\, J_m(|\bk_1+\bk_2|_{\alpha} r)\, J_m(|\bk_1+\bk_2|_{\alpha} r')\,.
\ee
Note that this integral involves only the Bessel functions of small order.
We now plug this expression in (\ref{sum2}) and get
\be
{1 \over 2\pi}\int_0^{2\pi} d\alpha\,
\left[\int rdr\, J_m(\kappa r)\, J_m(|\bk_1+\bk_2|_{\alpha} r)\right]
\left[\int r'dr'\, J_m(\kappa r')\, J_m(|\bk_1+\bk_2|_{\alpha} r')\right]
\ee
As usual, we extend the integration range in one of the integrals to infinity, which gives 
a delta-function, and then we use it on the second integral calculated up to $R$.
The original expression (\ref{integral-appendix}) then becomes
\be
{1 \over 2\pi} \int_0^{2\pi} d\alpha\,{\delta(\kappa-|\bk_1+\bk_2|_{\alpha}) \over \kappa} 
{R \over \pi \kappa} = {R \over 2\pi^2\kappa}\cdot {1\over \Delta}\,,\label{appendix-result}
\ee
where $\Delta$ is, as always, the area of the triangle with sides $\kappa$, $\kappa_1$, $\kappa_2$.

\end{document}